\documentclass[letter]{aa} 
\usepackage{graphicx}
\usepackage{txfonts}
\usepackage{lscape}
\usepackage{mathrsfs}
\usepackage{nicefrac}
\begin{document}
\title{Interior matter estimates of the X-ray pulsar in SAX J1808.4-3658 from mass-radius and rotation measurements}

\author{Nana Pan\inst{1}, Li Zhang \inst{1}, Xiaoping Zheng\inst{1}, }


\institute{The Institute of Astrophysics, Huazhong Normal
University, Wuhan 430079, Hubei, China\\
  \email{Pannana@phy.ccnu.edu.cn}\\
  \email{zhxp@phy.ccnu.edu.cn}
}

\authorrunning{Nana ~Pan et al.}

\titlerunning{ Interior matter estimates of the X-ray pulsar in
SAX J1808.4-3658}

\date{Received......; accepted......}


  \abstract
   {}
   {To constrain the equation of state of super-nuclear density matter
   and  probe the interior composition of the X-ray pulsar in SAX J1808.4-3658.
   In our estimation, we consider both its persistent 2.49 ms X-ray pulsations discovered by
   Wijnands and van der Klis from using the Rossi X-ray Timing Explorer,
   which is interpreted to come from an accreting-powered millisecond
   X-ray pulsar in the low mass X-ray
   binaries, and the corresponding mass-radius data analyzed of the light curves
   of SAX J1808.4-3685 during its 1998 and 2005 outbursts by Leahy
   et al. from assuming a hot spot model where the X-rays are
   originated from the surface of the neutron star.
    }
   {The interior composition of the X-ray pulsar in SAX J1808.4-3658 is obtained from comparing
   the mass-radius relations,
  Keplerian motions  and gravitational wave radiation instabilities of the predictions under different theoretical
   models with the analytically mass-radius district and rotation frequency of observational data.}
   {We show that SAX J1808.4-3658 may include the exotic matter
   such as hyperon or quark matter, but we can't distinguish the star with hyperon matter from quark matter. }
   {}


\maketitle
%

\section{Introduction}
The properties of nuclear matter under extreme conditions are
correlative to the in-depth experiment and theoretical research of
the heavy nuclear physics and nuclear astrophysics intimately.
Neutron stars provide us a unique spherical environment to study
the properties of matter with low temperature and high density.
Since the discovery of the first pulsar by Hewish et al. (1968)
and then the confirmation of it to be a fast rotational neutron
star by Gold and Pacini (1968), the compositions and properties of
the interiors of neutron stars have attracted much attention
(\cite{Pandharipande, Glendenning1, Glendenning2, Sahu, Thorsson,
Alford1}). If we have made this significant problem clear, it
would modify our understanding on the structure of the matter,
strengthen the cognition about the formation of the universe and
in all contribute most to the research of astrophysical physics,
nuclear physics and particle physics.

 However, in fact due to the uncertain nuclear physics, the
equation of state for neutron star at super-nuclear density is
still an indeterminacy, on which the maximum mass, radius and
 rotation frequency depend strongly. Many investigators
spontaneously expect to probe the matter under these conditions
and constrain or rule out some equations of state through
astrophysical observations (\cite{Glendenning3, Lattimer, Lackey,
Lavagetto, Klahn, Lattimer1, Pan}). Although researchers
universally impose the inferred masses and radii or rotation
frequencies for pulsars on the equations of state, their common
treatments cannot uniquely examine and distinguish all the classes
of  compositions inside neutron stars, i.e. it could rule out
neither traditional neutron stars nor hyperon stars or quark stars
determinately. This problem should be carefully considered in
other effective ways for the further estimation.

\section{The X-ray pulsar in SAX J1808.4-3658 }

In this letter, we mainly estimate the X-ray pulsar in the
transient X-ray burster SAX J1808.4-3658. Wijnands and van der
Klis discovered its persistent 2.49 ms X-ray pulsations from using
the Rossi X-ray Timing Explorer, and interpreted them to come from
an accreting-powered millisecond X-ray pulsar in the low mass
X-ray binaries (\cite{Wijnands}). Recently, Leahy
   et al. have analyzed its light curves during 1998 and 2005 outbursts
 to get the corresponding mass-radius data
 from assuming a hot spot model where the X-rays are
  originated from the surface of the neutron star (\cite{Leahy}).

\section{Probing stellar matter}
In what follows, we present our approach to probe the matter
composition of the X-ray pulsar in SAX J1808.4-3658 using the two
observational characteristics mentioned above.

\subsection{Mass-Radius estimates}
The structure of a neutron star is determined by the local balance
between the attractive gravitational force and the pressure force
of the neutron star matter. The X-ray pulsar in SAX J1808.4-3658
rotates at 2.49 ms, which in fact should be solved through the
perturbative approach of the non-rotating equilibrium
configuration in the framework of general relativity
(\cite{Hartle1,Hartle2}). But the rotation corrections of the
radii and masses are about 10\% and 2-3\%, which couldn't affect
our results significantly here, the non-rotating approximation
suffices.

 According to the
Tolman-Oppenheimer-Volkoff theory (TOV) under the consideration of
the effect of general relativity,
  the structures of spherically symmetric static neutron stars could be determined by a set of equations
  (\cite{Tolman,Opp}):
\begin{equation}
\frac{dp(r)}{dr}=-\frac{[\epsilon(r)+p(r)][m(r)+4 \pi
r^{3}p(r)]}{r[r-2m(r)]},
\end{equation}
\begin{equation}
\frac{dm(r)}{dr}=4\pi r^{2}\epsilon(r).
\end{equation}
($G=c=1$) where  $p(r)$ and $\epsilon(r)$ are the pressure and
energy density of the matter at the radius $r$, and $m(r)$ is the
total mass inside the star within a sphere of given radius $r$:
\begin{equation}
m(r)=4\pi \int_{0}^{r}\epsilon(r')r'^{2}dr'.
\end{equation}
After the equation of state of the star
$\epsilon(r)=\epsilon(p(r))$ is given, the TOV equations could be
 solved as an initial value problem. For a given equation of
 state, there exists a unique relationship between the stellar
 mass and radius. Thus different mass-radius relations obtained by considering
 various equations of state could be used to distinguish one from
 another.

\begin{table}
\caption{Summary of equations of state A1-D1 introduced in the
text, here compositions refers to the interacting components, n -
neutrons, p - protons,  e - electrons, H - hyperons and Q -
quarks. GC and MC represent the deconfinement phase transition
under Gibbs and Maxwell constructions respectively. RMF is the
abbreviation for relativistic field theoretical approach in the
mean field approximation, MIT the simple MIT bag model, eMIT the
effective mass MIT bag model, APR the equation of state of Akmal,
Pandharipande and Ravenhall using the variational chain summation,
pQCD perturbative QCD corrections, and BBG the
Brueckner-Bethe-Goldstone many-body approach. } \centering
\begin{tabular}{c|c|c|c}
\hline\hline  Symbols   &  Approaches &  References
&  Compositions \\
\hline
 A1  & RMF & Glendenning (1997)   & npe  \\
  A2  & RMF & Lackey et al. (2006)   & npeH  \\
  A3   & RMF+MIT & Pan $\&$ Zheng (2007)  & npeQ (GC) \\
   A4   & RMF+eMIT & Zheng et al. (2007)  & npeQ (MC) \\
    B1   & APR &  Akmal et al. (1998) & npe \\
     B(2-3)    & APR+pQCD &  Alford et al. (2005) & npeQ (GC/MC)\\
     C1   & BBG & Nicotra et al. (2006)   & npe  \\
      C(2-3)  & BBG+MIT & Nicotra et al. (2006)  & npeQ (GC/MC) \\
         & &  Fahri $\&$ Jaffe (1984)  &   \\
       C4  & BBG& Baldo et al. (2000)   & npeH  \\
       D1   & eMIT& Schertler et al. (1997)   & Q(u,d,s)  \\
\hline \hline
\end{tabular}
\label{tab:data}
\end{table}

As we have mentioned before, the neutron stars actually could have
many classes. Besides traditional neutron star (A1, B1, C1) that
mainly consists of neutrons (\cite{Glendenning, Akmal, Nicotra}),
when hyperons become populated in addition to the nucleons,  it
corresponds to the hyperon star (A2, C4) (\cite{Lackey, Baldo}).
In accordance with the predictions of quantum chromodynamics,
deconfined quark matter may also be produced there. When the
highly compressed matter could undergo the deconfinement phase
transition into u, d, s quark matter, or according to absolute
stable strange quark matter hypothesis, the quark-hybrid star
(A3-4, B2-3, C2-3) (\cite{Pan, Zheng, Alford, Nicotra, Fahri}) and
strange star (D1) (\cite{Schertler}) may exist. In Fig. 1, we
compare various radius-mass relations with the observation data of
SAX J1808.4-3658, of which the equations of state have been listed
detailedly in Table 1. Obviously, only these softer equations of
state could satisfy the inferred mass-radius region, and some of
the neutron stars, hyperon stars, quark stars and strange stars
match the observation well. But as predicted above, we can't tell
the exact components in the core of this neutron star.

\subsection{Rotation estimates}
For a uniform rigid compact star with mass M and radius R, there
exist some limits on the attainable rotation frequency. The most
obvious one is the Keplerian limit (\cite{Lattimer})
\begin{equation}
\nu_{K}=1.042\times10^{3}\times\frac{(M/M_{\bigodot})^{1/2}}{(R/10km)^{3/2}}
\textmd{Hz}, \label{KC}
\end{equation}
which is also called the mass-shedding limit. When its spin
frequency exceeds this limit, the matter near the equator would
run away, and the star may be no longer stable. It is foreign to
the material qualities of the matter inside. So if we get the
frequency of observed pulsar and assume it to be the Keplerian
limit, we can obtain a critical mass-radius relation to constrain
the equations of state, which has been given as the cyan district
in Fig. 1, but it also seems to have no help in distinguishing the
composition of the super-nuclear matter even if the mass-radius
method is considered together.

\begin{figure}
  \centering
  \includegraphics[width=0.51\textwidth]{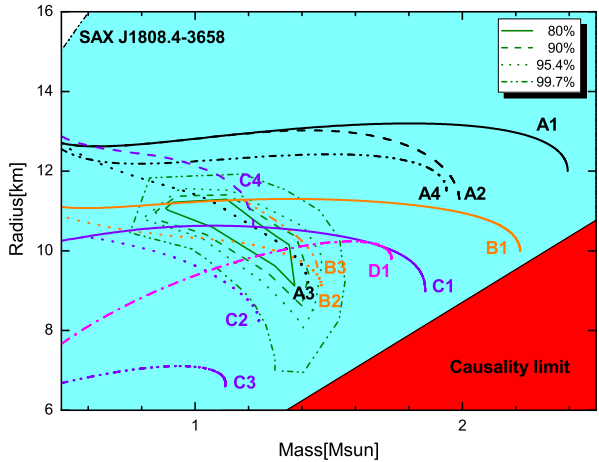}
  \caption{Radius-mass relations for various equations of state of compact stars listed in Table
  1. The red and cyan districts refer to the causality limit and
  simply traditional Keplerian rotation limit at 2.49 ms discovered by Wijnands and van der Klis (1998) respectively.
  These green contours correspond to 80\%, 90\%, 95.4\% and 99.7\% confidence levels of the X-ray pulsar
  in SAX J1808.4-3658 analyzed by Leahy et al. (2007). Note that
  we here use the non-rotating approximation in the calculation,
  as the radii and masses of these stars are about 10\% and 2-3\% larger due to the
  rotation correction at 2.49 ms, which couldn't affect our
  results significantly.
   }
    \label{mr}
\end{figure}

Actually, the emission of gravitational radiation following the
excitation of non-radial oscillation modes may lead to the
instability of rotating stars (\cite{Chandrasekhar, Andersson1,
Andersson}), and the corresponding limiting rotation can be
obtained via
\begin{equation}
\frac{1}{\tau_{G}}+\frac{1}{\tau_{\nu}}=0.
\end{equation}
Here $\tau_{G}<0$ is the characteristic time scale for energy loss
due to gravitational waves emission, and $\tau_{\nu}$ the damping
timescales due to shear, bulk viscosities and other rubbings,
which relates to the viscosities of the matter inside neutron
stars and their differences could result in diverse behaviors
(\cite{Pan2}).

In Fig. 2, we show the limit rotation frequencies for different
classes of neutron stars under the relativistic mean field theory
for demonstration, which is constructed by means of the method
brought forward by Zheng et al.(2007). Clearly, these thick solid
lines should be the genuine upper frequencies for each class.
After matching our theoretical predictions with the observation
data, we find that the traditional neutron star should be
excluded, although its Keplerian limit could reach above the 401
Hz of SAX J1808.4-3658. While hyperon star, quark star and strange
star are supposed to be the best candidates. Therefore, the star
 could contain either
hyperons or quarks at super-nuclear region, but we can't tell them
from each other. We should emphasize that, the conclusion is still
true if we apply it to various classes of neutron stars under
other approaches.

\section{Conclusions}
The composition  of matter in the core of neutron stars has
attracted much attention owing to its important significance. Up
to now, SAX J1808.4-3685 is one of seven known accreting ms X-ray
pulsar. Its persistent 2.49 ms X-ray pulsations was discovered by
Wijnands and van der Klis and interpreted as due to the ms
rotation of the center neutron star. And Leahy et al. have
analyzed its corresponding mass-radius relation. In our
estimation, we try to probe the inner components of it in our own
way by comparing the mass-radius relations and genuine rotation
frequencies under different theoretical models with the
observational data. We finally come to a conclusion that the
pulsar in SAX J1808.4-3685 is a star containing exotic matter,
which just bases on these two observational properties, but we
can't distinguish it with hyperon matter from quark matter at the
present time, which must depend on more observational information
about it, such as the thermal emission data.

\begin{figure}
\centering
  \includegraphics[width=0.51\textwidth]{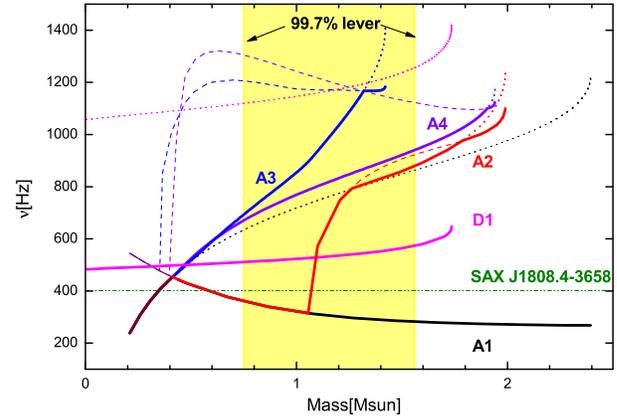}
  \caption{Limit rotation frequencies for various equations of state of compact stars. These dashed lines
represent the r-mode instability limits, and dotted lines the
Keplerian limits. The corresponding thick solid lines are the
genuine upper frequencies. The light yellow district refers to the
mass prediction of the X-ray pulsar in SAX J1808.4-3658 at 99.7\%
confidence lever analyzed by Leahy et al. (2007), and the green
dash-dot-dot horizontal line the corresponding 2.49 ms rotation
frequency discovered by Wijnands and van der Klis (1998). }
\label{mf}
\end{figure}

\begin{acknowledgements}

\end{acknowledgements}
This work is supported by the National Natural Science Foundation
of China under Grant Nos. 10773004 and 10603002.

\end{document}